\documentstyle[sprocl,epsfig]{article}

\def\be{\begin{equation}}
\def\ee{\end{equation}}
\def\bea{\begin{eqnarray}}
\def\eea{\end{eqnarray}}

\begin{document}

\vspace*{-9\baselineskip}
\begin{flushright}
\hfill {\rm UR-1598}\\
\hfill {\rm ER/40685/943}\\
\hfill {\rm October 1999}\\
\end{flushright}
\vspace*{2\baselineskip}

\title{GLUON RADIATION IN TOP PRODUCTION AND DECAY\footnote{Presented at the 
International Workshop on Linear Colliders, Sitges, Spain, April 28-May 5, 1999}}

\author{ COSMIN MACESANU, LYNNE H.~ORR }

\address{Department of Physics and Astronomy, University of Rochester,
Rochester, NY 60627-0171, USA}


\maketitle\abstracts{We present the results of an exact calculation of gluon 
radiation in top production and decay at high energy electron-positron
colliders.  We include all spin correlations and interferences,  the bottom
quark mass, and  finite top width  effects in the matrix element
calculation. 
We study properties of the radiated gluons and implications for top 
mass measurement.
}
  
\section{Monte Carlo Calculation}

A high energy linear electron-positron collider will provide an excellent 
environment for producing and studying the top quark, free of the 
large QCD backgrounds produced in hadron colliders.  An important QCD
effect which does occur at $e^+e^-$ colliders, however, is the radiation
of gluons from the produced top quark and/or its decay products.  
Radiated gluons appear as additional jets in top events which can
complicate event identification and top mass reconstruction.

In this talk we
examine gluon radiation in the production  and decay ($t\to Wb$) of top 
quark pairs at 
high energies\cite{mo}; see also \cite{schmidt}.  At lepton 
colliders, there is no gluon radiation from the initial state, but 
the top quarks and their daughter $b$ quarks can emit gluons.\footnote{We
neglect radiation from decay products of the $W$ bosons.}  
We perform a Monte Carlo calculation of the process 
\begin{equation}
e^+e^- \rightarrow \gamma^*, Z^* \rightarrow t\bar{t} (g)
\rightarrow bW^+ \bar{b}W^-g\; .
\end{equation}
We compute exact matrix elements of the contributing diagrams 
with all spin correlations and the bottom mass
included.  We keep the 
finite top width $\Gamma_t$ in the top quark propagator and include
all interferences between diagrams, and we use exact kinematics in
all parts of the calculation.

Radiation
from the top quarks can occur as part of the top production process
(before the top quarks go on shell) or as part of the decay process
(after the top go on shell).   Radiation from the $b$ quarks is always
part of top decay.   For reconstructing the top quark mass (to
identify top events and to measure $m_t$), whether to include the gluon
in the reconstruction is determined by whether the 
radiated gluon is part of the production or decay process.  If it is
emitted in the production stage, we obtain the top mass from the $W$ and 
$b$ momenta only ($m_t^2=p_{Wb}^2$), but if it is part of the decay, we
must include it in the reconstruction ($m_t^2=p_{Wbg}^2$).

In practice this distinction cannot be made absolutely in an experiment,
but it is useful to make it in the calculation.  There is a single 
Feynman diagram that contributes to both production- and decay-
stage radiation from the top quark, and the corresponding 
matrix element contains two propagators:
\begin{equation}
ME \propto \left( {{1}\over{p_{Wbg}^2-m_t^2+im_t\Gamma_t}}\right)
\left( {{1}\over{p_{Wb}^2-m_t^2+im_t\Gamma_t}}\right)\; .
\end{equation}
We can rewrite the product as 
\begin{equation}
{{1}\over{2p_{Wb}*p_{Wbg}}}
\left( {{1}\over{p_{Wb}^2-m_t^2+im_t\Gamma_t}} -
{{1}\over{p_{Wbg}^2-m_t^2+im_t\Gamma_t}}\right)\; ,
\end{equation}
which separates the production and decay contributions according to
where the terms in parentheses peak.  The cross section in 
turn contains separate production and decay contributions.\footnote{It also 
contains
interference terms, which in principle confound the separation but in 
practice are  small.}

\section{Results}

\begin{figure*}[t]
\vskip -4cm
\psfig{figure=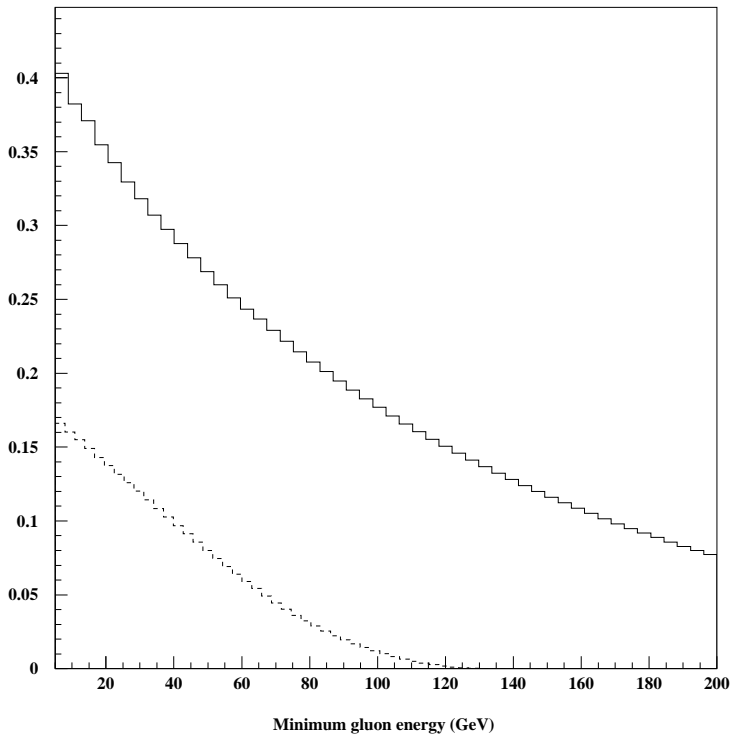,width=11cm}
\vskip -.5cm
\caption[]{  
The fraction of gluon emissions radiated in the production stage, as
a function of minimu gluon energy, for center-of-mass energy 1 TeV (solid
line) and 500 GeV (dashed line).\label{prodfrac}}
\end{figure*}

We show in Figure \ref{prodfrac} the fraction of the total cross section due 
to production stage 
emission, in events with an extra gluon, as a function of the minimum gluon 
energy.  The solid histogram is for center of mass energy 1 TeV, and the 
dashed histogram for 500 GeV.  The main effect is due to kinematics --- both 
curves fall off with increasing gluon energy because of a reduction in 
the phase space for gluon emission, and the production fraction is always 
smaller for the lower collision energy.  In both cases decay-stage radiaion 
dominates for all gluon energy thresholds.

\begin{figure*}[t]
\vskip -4cm
\psfig{figure=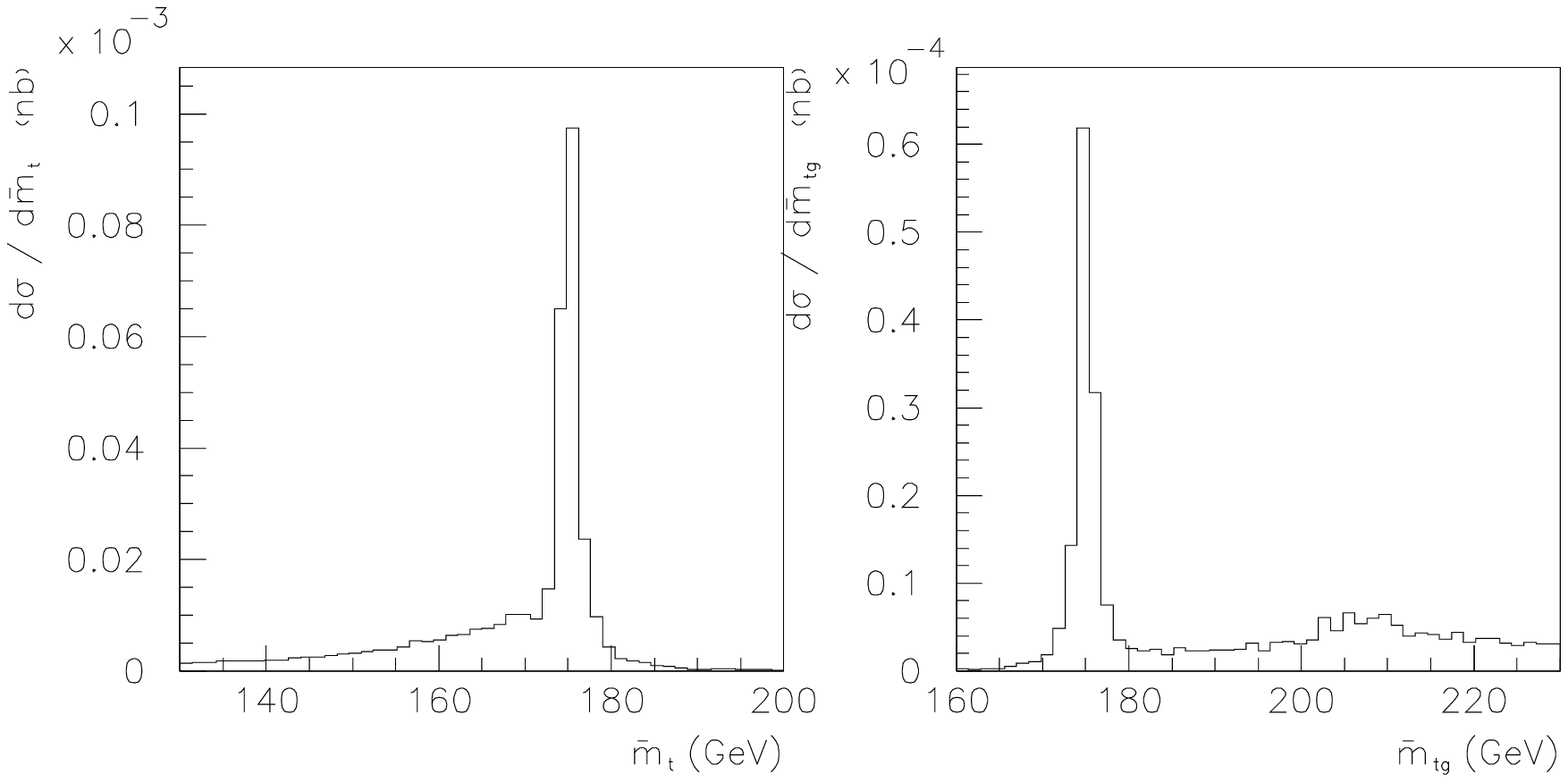,width=11cm}
\vskip -.75cm
\caption[]{  The  top invariant mass spectrum without (left) and with (right) 
the gluon momentum included, for center-of-mass energy 600 GeV.
\label{masses}}
\end{figure*}

We now consider  mass reconstruction.  
Figure \ref{masses} shows top invariant mass distributions with and without
the extra gluon included. In both cases there is a clear peak  at the 
correct value of $m_t$.  In the left-hand plot, where the gluon
is not included in the reconstruction, we see a low-side tail due to 
events where the gluon was radiated in the decay.  Similarly, 
in the right-hand plot we see a high-side tail due to events where the 
gluon was radiated in association with production, and was included when it 
should not have been.

The narrowness of the peaks and the length of the tails in Figure \ref{masses}
suggests that an invariant mass cut would be useful to separate the two
types of events.  We can in fact do even better using cuts on  
the angle 
between the gluon and the $b$ quarks.    This works because 
although there is no collinear singularity for radiation from massive
quarks, the distribution of gluons radiated from $b$ quarks peaks
close to the $b$ direction.  Such gluons are emitted in decays.
We should note that an important reason the cuts are so effective is 
that we work at the 
parton level.  In an experiment, hadronization and detector effects are 
likely to significantly reduce the resolution.

Finally we turn to the interference between the production- and decay-stage 
radiation, which though small is interesting for its sensitivity to 
the top width $\Gamma_t$ (about 1.5 GeV in the Standard 
Model).\cite{kos}
The interference between the two propagators shown above can be 
thought of as giving rise to two overlapping Breit-Wigner resonances.  The
peaks are separated roughly by the gluon energy, and each curve
has width $\Gamma_t$.  Therefore when the gluon energy becomes comparable 
to the top
width, the two Breit-Wigners overlap and interference can be substantial.
In constrast, if the gluon energy is much larger than $\Gamma_t$, overlap
and hence interference is negligible.  Hence the amount of interference serves
as a measure of the top width.  
 
\begin{figure*}[t]
\vskip -4cm
\psfig{figure=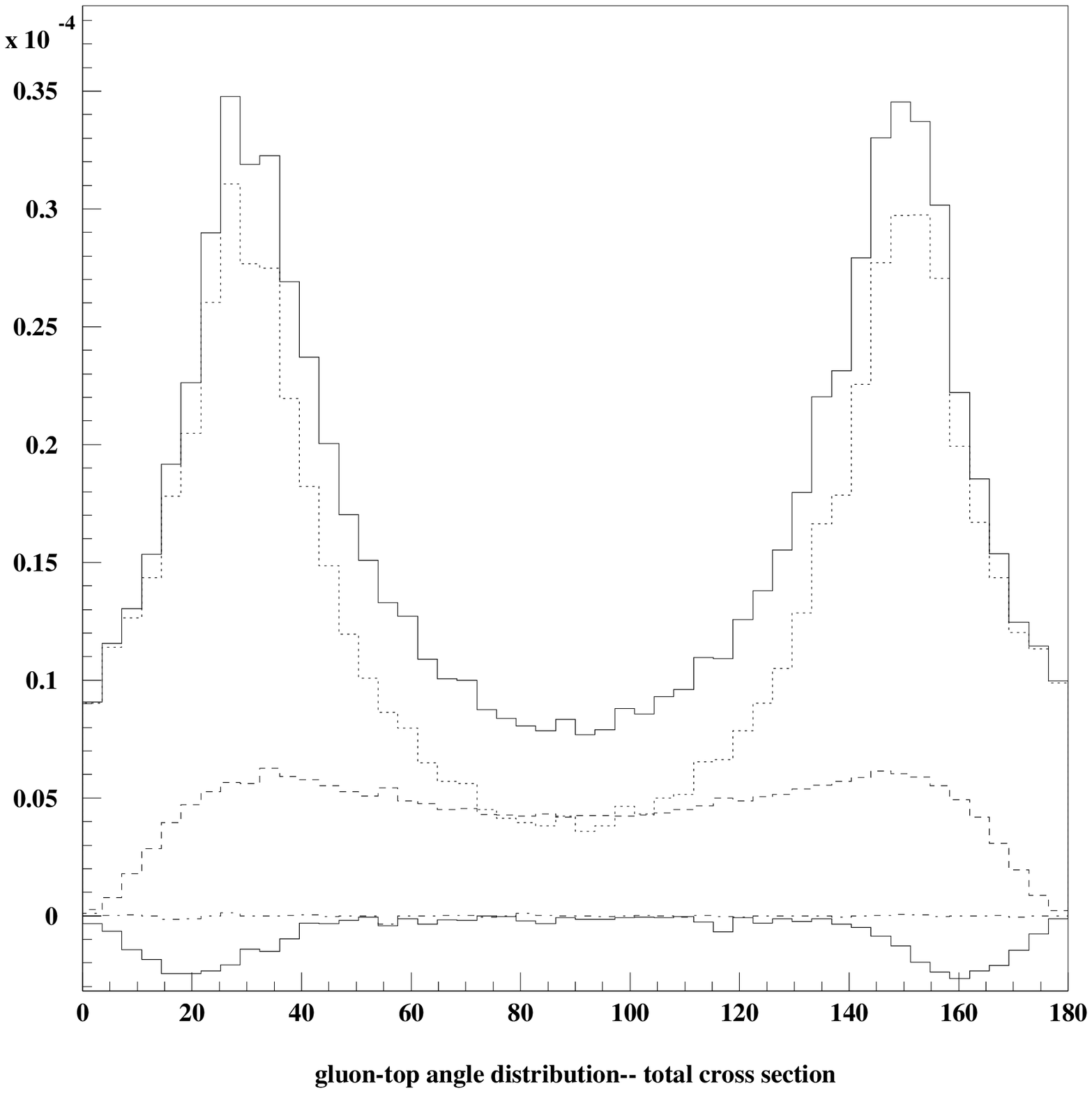,width=11cm}
\vskip -.5cm
\caption[]{The  distribution in angle between the top quark and the gluon for 
gluon energies from 5 to 10 GeV, $\cos\theta_{tb}<0.9$, and 750 GeV collision
energy.  The upper solid histogram is the  total and the other histograms
represent the individual contributions:  dotted:  decay; dashed: production;
dot-dashed: decay-decay interference; solid:  production-decay interference.  
\label{thetagt}}
\end{figure*}

This effect was discussed in \cite{kos} using the soft gluon approximation,
and Figure \ref{thetagt} shows that the effect remains in the exact calculation.
There we plot the distribution in the angle between the 
emitted gluon and the top quark for gluon energies between 5 and 10 
GeV and with $\cos\theta_{tb}<0.9$.  The center-of-mass energy is 750 GeV.
The  histograms show the decomposition into the various contributions.  
The negative solid histogram is the production-decay interference, and 
we see that not only is it substantial, it is also destructive.  That means
that the interference serves to suppress the cross section.  If the 
top width is increased, the interference is larger, further suppressing the 
cross section.  Although
the sensitivity does not suggest a precision measurement, it is worth
noting that the top width is difficult to measure by any means, and it
is the total width that appears here. 

In summary, we have presented preliminary results from an exact
parton-level calculation of real gluon radiation in top production and decay
at lepton colliders, with the $b$ quark mass and finite top width, as well
as all spin correlations and interferences included.  We have indicated
some of the issues associated with this gluon radiation in top mass
reconstruction and top width sensitivity in the gluon distribution.  Further
work is in progress.

\section*{Acknowledgments}
Work supported in part by the U.S. Department of Energy,
under grant DE-FG02-91ER40685 and by the U.S. National Science Foundation, 
under grant PHY-9600155.


\section*{References}
\begin{thebibliography}{99}

\bibitem{mo} C.~Macesanu and L.H.~Orr, in preparation.
\bibitem{schmidt} C.R.~Schmidt, Phys.~Rev.\ 
{\bf D54} 3250 (1996). 
\bibitem{kos} V.~Khoze, L.H.~Orr and  W.J.~Stirling, 
Nucl.~Phys.  {\bf B378} 413 (1992). 


\end{thebibliography}
\end{document}